# Gate-tuneable and chirality-dependent charge-to-spin conversion in Tellurium nanowires

Francesco CALAVALLE[1,†], Manuel SUÁREZ-RODRÍGUEZ[1,†], Beatriz MARTÍN-GARCÍA[1], Annika JOHANSSON[2,3], Diogo C. VAZ[1], Haozhe YANG[1], Igor V. MAZNICHENKO[2], Sergey OSTANIN[2] Aurelio MATEO-ALONSO[4,5], Andrey CHUVILIN[1,4], Ingrid MERTIG[2], Marco GOBBI[1,4,6,*], Fèlix CASANOVA[1,4,*] and Luis E. HUESO[1,4,*]

[1]CIC nanoGUNE BRTA, 20018 Donostia-San Sebastian, Basque Country, Spain

[2] Institute of Physics, Martin Luther University Halle-Wittenberg, 06099 Halle, Germany

[3] Max Planck Institute of Microstructure Physics, Weinberg 2, 06120 Halle, Germany

[4]IKERBASQUE, Basque Foundation for Science, 48013 Bilbao, Basque Country, Spain

[5]POLYMAT, University of the Basque Country UPV/EHU, 20018 Donostia-San Sebastian, Basque Country, Spain

[6]Centro de Física de Materiales CSIC-UPV/EHU, 20018 Donostia-San Sebastian, Basque Country, Spain

* Correspondence to: m.gobbi@nanogune.eu; f.casanova@nanogune.eu; l.hueso@nanogune.eu

† These authors contributed equally to this work

Abstract

Chiral materials are the ideal playground for exploring the relation between symmetry, relativistic effects, and electronic transport. For instance, chiral organic molecules have been intensively studied to electrically generate spin-polarized currents in the last decade, but their poor electronic conductivity limits their potential for applications. Conversely, chiral inorganic materials such as Tellurium are excellent electrical transport materials, but have not been explored to enable the electrical control of spin polarization in devices. Here, we demonstrate the all-electrical generation, manipulation, and detection of spin polarization in chiral single-crystalline Tellurium nanowires. By recording a large (up to 7%) and chirality-dependent unidirectional magnetoresistance, we show that the orientation of the electrically generated spin polarization is determined by the nanowire handedness and uniquely follows the current direction, while its magnitude can be manipulated by an electrostatic gate. Our results pave the way for the development of magnet-free chirality-based spintronic devices.

Introduction

Charge-to-spin interconversion phenomena such as the spin Hall effect[1] or the Edelstein effect[2] enable the electrical generation of spin currents without magnetic elements[3], a fundamental step towards the next generation of spintronic devices, such as spin-based logic[4,5] and spin-orbit torque[6] MRAM memories[7]. The Edelstein effect[2], also called inverse spin galvanic effect[8], emerges in materials with strong spin-orbit coupling (SOC) and broken inversion symmetry, such as strained semiconductors[9], Rashba systems[10–12] and the surface of topological insulators (TIs)[13,14]. In these systems, the spins of the electrons are locked in the direction perpendicular to their momenta, so that the flow of a charge current results in a perpendicularly oriented homogeneous spin polarization. Further lowering the crystal symmetry allows the creation of spin polarizations in unconventional directions[15,16] and enables new fundamental effects[17,18] and configurations for devices[19–24].

Chiral materials are the ultimate expression of broken symmetry, lacking inversion and mirror symmetry. Indeed, the relation between structural chirality and spin effects has been investigated for organic molecules[25–28], which act as efficient spin filters in spite of their low electrical conductivity. This phenomenon, which is often named chiral-induced spin selectivity (CISS), remains poorly explored in chiral inorganic crystals [29–32]. Tellurium (Te), a material that possesses strong SOC, a chiral structure[33,34] and can be synthesized[35] in nanowires (NWs) or flakes with excellent electronic conductivity[36,37], is an ideal material for the study of unconventional chirality-related charge-to-spin conversion. Signs of current-induced spin polarization have been provided only in bulk Te through optical[38,39] and NMR measurements[40,41], but these detection techniques and the use of millimetre-sized crystals are not suitable for integration into all-electrical nanodevices.

Here, we report a chirality-dependent and gate tuneable Edelstein effect in naturally hole-doped single-crystalline Te NWs. A net spin polarization is detected by recording a unidirectional magnetoresistance (UMR)[11,42–45] dependent on the relative orientation of the electrical current and the external applied magnetic field. Our results show that, unlike Rashba systems and TIs, the charge current flow in Te leads to a spin polarization oriented along the current path and pointing in opposite directions for left- or right-handed NWs. The measured UMR is explained on the basis of a chirality dependent Edelstein effect arising from the radial spin texture at the *H*-point of the valence band of Te, which dominates the transport in our hole-doped Te NWs. Using a conventional UMR figure of merit[45], we show that this effect in Te is the largest yet observed. More importantly, the electrostatic gating of the Te NWs allows us to tune its Edelstein effect leading to an electrical tuning of the UMR amplitude by a factor of 6. The all-electrical generation, control, and detection of spin polarization in chiral Te NWs opens the path to exploit chirality in the design of novel solid-state spintronic devices.

Results

A hydrothermal process in the presence of a reducing agent was employed to grow single crystalline Te NWs which are tens-of-micrometres long and a few hundred nanometres wide (see Methods section and Fig. 1a)[37]. Scanning transmission electron microscopy (STEM) was employed to confirm that the Te NWs are single crystals displaying one of the two enantiomorphic chiral space groups, $P3_121$ and $P3_221$ (ref. [33,34]). Figure 1b displays a 3D sketch of the right- and left-handed Te chains arranged in their characteristic helical structure that twists around the chiral c axis. Van der Waals forces keep together adjacent atomic chains, which possess the same helicity, providing a defined chirality to the Te crystal structure. Figure 1c shows a STEM image of a NW cross section perpendicular to the *c*-axis. Individual chains appear as triangles stemming from the projection of superimposed Te atoms. This atomic arrangement (see Fig. 1b) indicates that the long axis of the Te NWs is oriented along the chain direction (*c*-axis). While the image in Fig. 1c shows the high quality of the NW (see also Supplementary Section 1), it cannot be used to identify its chirality. However, it is possible to unambiguously distinguish between right- and left-handed crystals through a comparison between the atomic arrangement in different crystalline planes[33,34]. In Fig. 1d, we observe that the STEM images of two NWs with opposite handedness match with the models for the space groups $P3_121$ and $P3_221$.

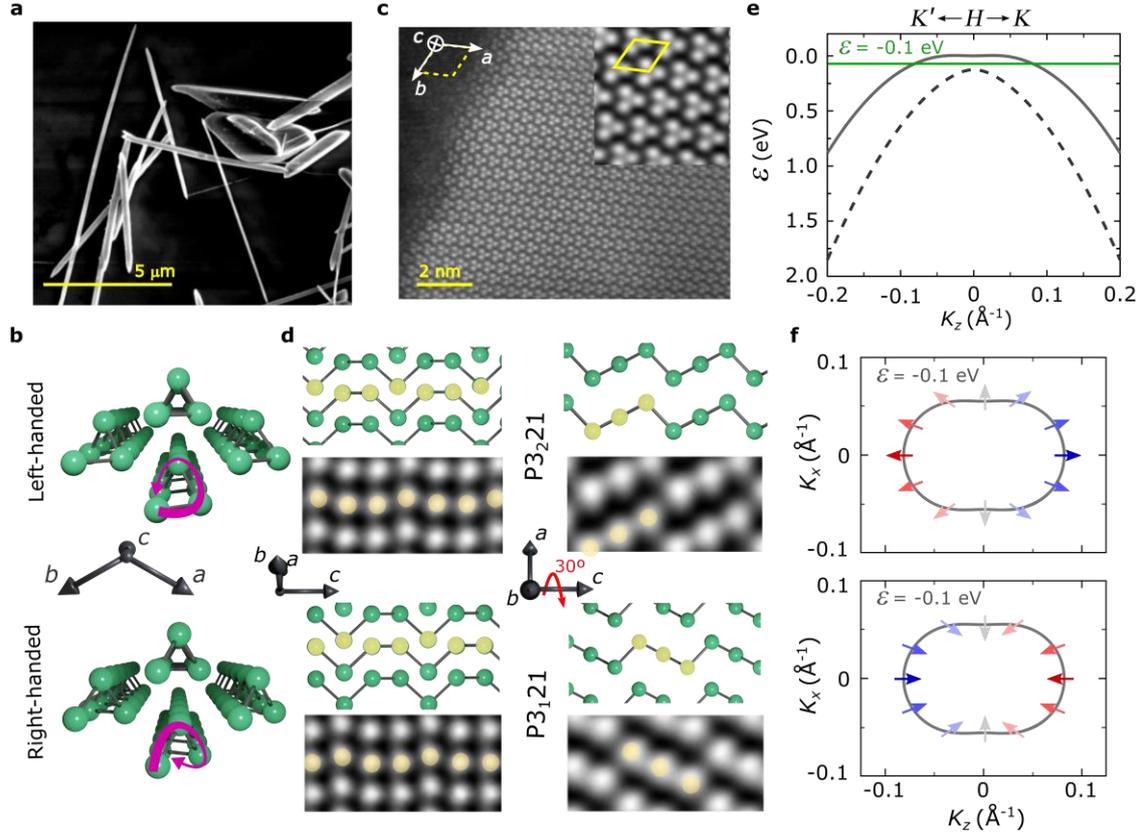

**Fig. 1| Crystallographic characterization, electronic band structure and spin texture of Te nanowires. a**, SEM image of Te NWs drop-casted on a TEM grid. **b**, 3D sketches of the right- and left-handed crystal structure of trigonal Te. **c**, STEM image of a Te lamella, obtained cutting the NW perpendicularly to the *c*-axis. **d**, Crystal structure sketch and STEM image of two Te NWs with opposite chirality, imaged at [−110] orientation and after a 30º rotation around the *c*-axis ([010]). **e**, Te band structure (valence bands) around the H-point, obtained from an effective model Hamiltonian[39] (see Methods) with SOC included. The solid and dashed lines represent respectively the upper and lower valence band around the H-point. **f**, Iso-energy contours of right- and left-handed Te at $\varepsilon = -0.1$ eV, showing the characteristic radial spin texture. The red/blue colours illustrate the (+/−) *z*-component of the expectation value of the spin operator.

The lack of mirror and inversion symmetries, deriving from the chiral nature of Te, in combination with a high SOC, result in a complex band structure characterized by non-degenerate spin bands at highly symmetric *k*-points (for more details on the band structure see Supplementary Section 2). The relevant region for electrical transport is located around the *H*-point, where a bandgap of 0.335 eV separates the conduction and valence band, and where the two upper branches of the valence band are separated by 0.126 eV (Ref. [39]). Figure 1e shows a zoom of the valence bands around this point calculated from an effective model (see Methods)[39]. The iso-energy contour corresponding to a cut of the valence band at $\varepsilon = -0.1$ eV below the band edge is characterized by a radial spin texture with spins pointing inward/outward in right-/left-handed crystals, as shown in Fig. 1f. Recent spectroscopic observations confirmed the link between the handedness of Te and the direction of the radial spin texture[46,47]. This is analogous to spin momentum locking observed in bulk Rashba systems[11,12], but with a chirality-dependent radial spin texture instead of the conventional helical configuration[3].

For the electrical characterization, we transferred Te NWs onto Si/SiO$_2$ substrates using a Langmuir-Schaefer approach. Individual NWs were subsequently selected and contacted using Pt contacts defined by standard lithography methods (see Methods). Figure 2a shows the optical image of a contacted Te NW, with a sketch of the four-probe configuration used for the transport experiments. We redefine a cartesian coordinate system to describe the directions of currents and magnetic fields applied, where z is in the same direction as the *c* crystallographic axis and *x*(*y*) are orthogonal to *z* and directed in(out) of the device plane (see Supplementary Section 3 for more details). Figures 2b-d show the magnetotransport characterization of a typical Te NW. For these measurements, we plot the average resistance $R^{avg} = [R(+I_z) + R(-I_z)]/2$, where $R(+I_z)$ and $R(-I_z)$ are the resistances measured with a d.c. positive and negative current, respectively (see Methods). This way, we obtain the equivalent of the 1$^{st}$ harmonic response in a.c. transport measurements[44], excluding current-dependent contributions to the resistance. The temperature (*T*) dependence of the four-probe resistance shows a monotonic decrease of *R* with decreasing T (Fig. 2b). From the transfer characteristics, we observed that the Te NWs are hole-doped, with a field effect mobility that ranges from ~ 500 cm$^2$/Vs at 300 K to ~ 2500 cm$^2$/Vs at 10 K (see Supplementary Section 4). This is in agreement with previous reports highlighting that Te vacancies cause hole-doping[48]. As a consequence of this intrinsic doping, the resistivity at room temperature is $\rho = 0.02$ Ω cm, one order of magnitude lower than the value reported for undoped bulk crystals[49].

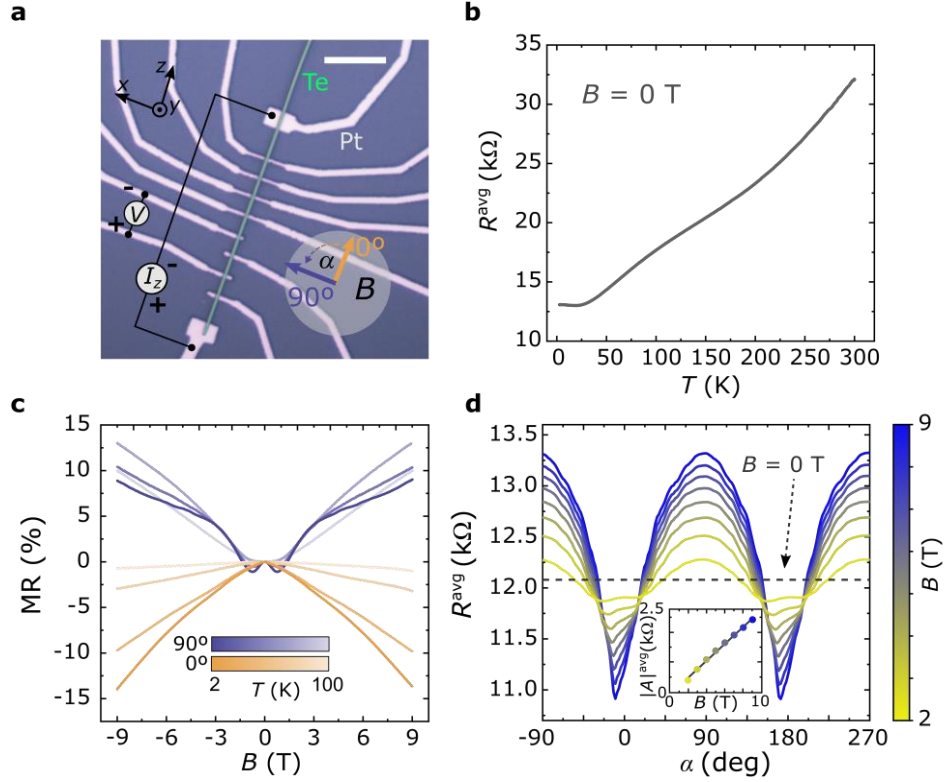

**Fig. 2| Magnetoelectrical characterization of a Te NW. a**, Optical image of a typical Te NW contacted with Pt contacts (the scale bar corresponds to 10 μm), illustrating the scheme of the four-probe measurement configuration. The angle α defining the orientation between the magnetic field and the NW is also drawn. **b**, Temperature dependence of the four-probe resistance $R^{avg} = [R(+I_z) + R(-I_z)]/2$ at zero magnetic field. **c**, Te Magnetoresistance (MR = $[R^{avg}(B) - R^{avg}(0)]/R^{avg}(0)$) measured along (α = 0º) and transversally (α = 90º) to the chiral z-axis at different temperatures (2, 10, 50 and 100 K). **d,** Angular dependence of $R^{avg}$ at different magnetic fields and $T$ = 10 K. The dashed line indicates the resistance at zero field. Small shifts in angle are caused by misalignments introduced when mounting the sample on the chip carrier. Inset: the variation of the angular-dependent MR amplitude ($|A^{avg}| = |R^{avg}(90º) - R^{avg}(0º)|$) with the field applied.

Figure 2c displays the magnetoresistance curves for in-plane magnetic fields $B$ parallel (0º) or perpendicular (90º) to the applied current $I_z$, at temperatures ranging from 2 to 100 K. When $B$ is parallel to $I_z$ (orange curves in Fig. 2c), we observe a monotonical decrease of the resistance when increasing the magnetic field. The negative longitudinal magnetoresistance (NLMR), which is unusual for non-magnetic materials, reaches a maximum value of −15% at $T$ = 2 K and $B$ = 9 T. This peculiar behaviour was already reported for bulk Te[48], and it has been attributed to the presence of Weyl Fermions contributing to the transport, although other effects may also be responsible for such MR (see supplementary section 5).

In the case of $B$ perpendicular to $I_z$ (purple curves in Fig. 2c), the magnetoresistance at low $T$ changes from negative to positive when increasing the field, showing a relative maximum at $B$ = 0 T and two minima around $B$ = ± 1 T, followed by a change of slope at around $B$ = 3 T. The minima slowly flatten out with increasing $T$, up to $T$ = 50 K, where they disappear (see section 5 of SI for more details.). In Fig. 2d, we show the angular dependence of the magnetoresistance at $T$ = 10 K as a function of different magnetic fields. Here and in the following experiments, we set the angle between the magnetic field and the NWs to be $α$ = 90º, when the field is orthogonal to the NW's $z$-axis and to the current applied, and at $α$ = 0º, when the field is parallel to the NW's $z$-axis and to the positive current direction (see Fig. 2a). The curves are characterized by sharp minima at around $α$ = 0º and $α$ = 180º, i.e., for $B$ // $I_z$, and maxima at around $α$ = ±90º, for $B$ ⊥ $I_z$. We highlight that, for our study, we focused our attention on the regime at $B$ > 3 T where, for every field, the resistance measured at ±90º (0º and 180º) is higher (lower) than the resistance measured at $B$ = 0 T, which is shown as a dashed line. The appearance of the maxima and minima is a direct consequence of the MR curves shown in Fig. 2c, which transit from positive at 90º to negative at 0º. Additionally, we notice that, at 10 K and for $B$ > 3 T, the dependence of the resistance on the field is almost linear for both the transverse and the longitudinal MR, but with an opposite slope (Fig. 2c). Consequently, the difference between the maximum and the minimum value in the angular dependence ($|A^{avg}|$) increases linearly with the applied field (see inset in Fig. 2d).

We now focus on the detection of the chirality-dependent current-induced spin polarization in the Te NWs. Figure 3a,b displays a sketch of the radial spin texture of Te for left- and right-handed crystals. An electric field applied along the $z$-direction generates a current density $j_z$ and causes a redistribution of states along the $k_z$ direction, which is depicted as a $\Delta k_z$ shift of the Fermi contour. Due to the radial spin texture of Te, $\Delta k_z$ induces a homogeneous spin density that is parallel or antiparallel to $j_z$, depending on the chirality of the NWs. Unlike Rashba systems, the electrical current in Te is expected to acquire a net spin polarization oriented along $j_z$, which is a peculiar manifestation of the Te symmetries (see Supplementary Section 6). This current induced spin polarization can be detected in magnetotransport measurements since it introduces a dependence of the resistance on the mutual orientation of current and magnetic field. This effect, also known as UMR or non-reciprocal charge transport, allows to map the spin texture of a material[11,42,44,45].

Using the configurations of current and magnetic field illustrated in Fig. 3c, we analysed the angle-dependent magnetoresistance of the NWs, measured for opposite current directions along $z$ ($\pm I_z$). By changing the current direction and varying the angle α between the NWs at a fixed magnetic field, we reverse the mutual orientation of $I_z$ and $B$. A parallel alignment between field and current is obtained at 0° for $+I_z$ and at 180° for $-I_z$ (see Fig. 3c).

Figures 3d,e show the resistance measured for $+I_z$ and $-I_z$ as a function of the angle α between the NW and a magnetic field ($B = 9$ T), for the two NWs with opposite chirality (cross-sections of these two same devices are shown in Fig. 1d). For both the right- and left-handed NW, we find a strong dependence of the resistance on the relative alignment between $I_z$ and $B$, which is the hallmark of UMR, and it demonstrates the presence of a net spin polarization in the Te NWs. In particular, the left-handed NW shows a significantly higher (lower) resistance when the current is parallel (antiparallel) to the external magnetic field (Fig. 3d). This non-reciprocal effect can also be observed as a shift in the MR traces measured for opposite current directions (see Supplementary Section 5, Fig. S5). Conversely, the right-handed NW displays lower (higher) resistance for $I_z$ parallel (antiparallel) to $B$ (Fig. 3e), indicating that the UMR of Te is mirrored in NWs with opposite chirality. Hence, the current induced spin polarization is reversed for opposite NW handedness.

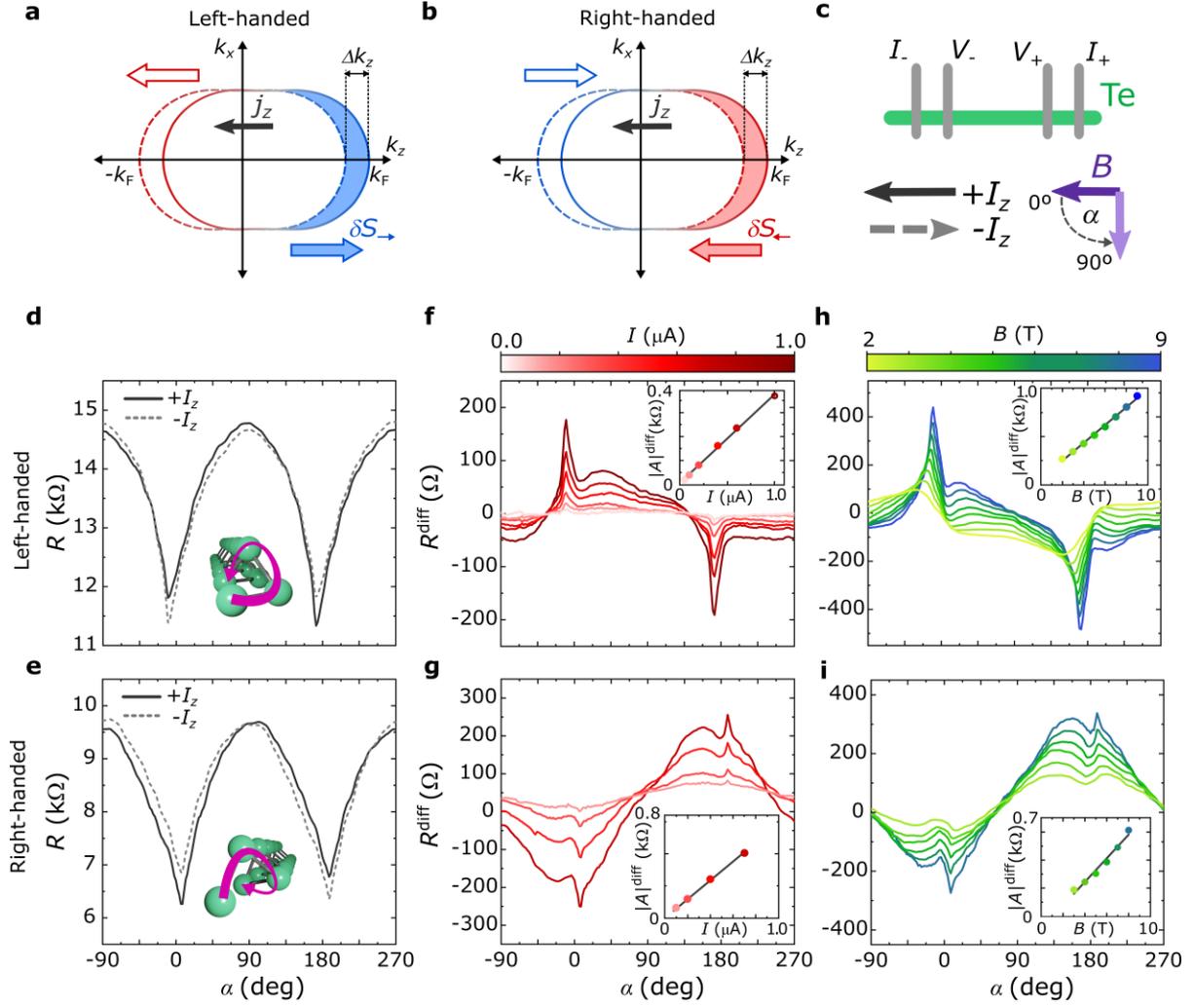

**Fig. 3| Unidirectional Magnetoresistance in a right- and a left-handed Te nanowire. a,b** Sketch representing the Edelstein mechanism responsible for the UMR. The shift in $k_z$ of the Fermi contours translates in the formation of spin densities oriented in opposite direction, due to the chirality-dependent spin texture. Empty arrows mean depletion of states, filled arrows mean higher occupation of states. **c**, Schematic of the sample and the measurement configurations defining the angle of the magnetic field with respect to the current directions. **d,e** Angular dependences of the magnetoresistance measured at 9 T and 10 K for two Te NWs with opposites handedness (confirmed by TEM analysis). Solid and dashed lines indicate the signal obtained from opposite current directions ($\pm I_z = \pm 1$ μA in **d** and $\pm I_z = \pm 0.7$ μA in **e**). **f-i**, Angle-dependent UMR as a function of the applied current (**f,g**, $B = 9$ T); and as function of the magnetic fields (**h**, $\pm I_z = \pm 5$ μA and **i**, $\pm I_z = \pm 1$ μA). The signal $R^{\text{diff}} = [R(+I_z) - R(-I_z)]/2$ shows a UMR with specular features between left- (**f,h**) and right- (**g,i**) handed NWs. In the insets, the amplitude $|A|^{\text{diff}} = |R(0°) - R(180°)|$ is represented as a function of magnetic field and current, showing in both cases a linear behaviour.

To analyse these data, we calculate the half difference between the resistance measured applying $+I_z$ and $-I_z$ ($R^{\text{diff}} = [R(+I_z) - R(-I_z)]/2$) obtaining the equivalent of the 2$^{\text{nd}}$ harmonic signal in a.c. transport measurements[44], providing a direct signature of UMR. In Fig. 3f-i, the angular dependence $R^{\text{diff}}(\alpha)$ presents sharp peaks for collinear current and magnetic field (at 0º and 180º), corresponding to the asymmetries in Fig. 3d,e. This indicates that, in agreement with the Edelstein mechanism (Fig. 3a,b), the spin polarization is oriented along the direction of the current, parallel to the chiral axis z. The deviation of $R^{\text{diff}}(\alpha)$ from the sin dependence typically observed for UMR signals can be explained by analysing the presence of additional UMR components not related to the chirality and the dependence of UMR on the peculiar MR recorded in our Te NWs (see Supplementary Section 5).

Moreover, the trend of the magnetoresponse (measured in right- and left-handed NWs) is specular since it is directly related to the chirality of the NWs. In this regard, the orientation of spin polarization is determined by the Te handedness, in a similar way the spin polarization in a ferromagnetic metal is determined by its magnetization. By matching the transport measurements with the STEM analysis in Fig. 1, we can distinguish the NWs handedness through their magnetotransport behaviour.

Figure 3f,g shows the variation of $R^{\text{diff}}(\alpha)$ measured at $B = 9$ T and different current $I_z$. A larger magnetoresponse is measured at higher currents, with the signal amplitude $|A|^{\text{diff}} = |R(0º) - R(180º)|$ increasing linearly with $I$ (see insets). This dependence can be understood considering that the spin density generated by the Edelstein effect increases linearly with the displacement $\Delta k_z$ induced by the current. According to the model in Fig. 3a,b, a higher current induces a larger spin density, resulting in more spins coupling to the magnetic field.

Figure 3h,i shows $R^{\text{diff}}(\alpha)$ measured for fixed current and different magnetic fields. Even in this case, we observe a linear increase of $|A|^{\text{diff}}$ with $B$, which indicates that the UMR recorded in our Te NWs possesses a bilinear response to current and magnetic field. This bilinearity, which has been observed in spin-momentum locked states in Rashba systems[42] and in TIs[44], originates from the $B$ dependent relaxation processes at scalar and spin-orbit defects, respectively[42,43].

Besides the fundamental connection between the chirality of Te and its magnetotransport, we highlight that the UMR amplitude in our NWs is very substantial. For the NW shown in Fig. 3d,f,h, the maximum $|A|^{\text{diff}}$ measured at 9 T and 5 µA amounts to 800 Ω, which corresponds to a 7% variation with respect to the current-averaged resistance measured at the same field and at the same angle, $R^{\text{avg}}(0°, 9\text{ T}) = 10.8$ kΩ. To the best of our knowledge, in previous reports of UMR, the relative variation in resistance was always < 1% (Ref. [11,42,44,45,50]). To further compare our results with previous works, we use the figure of merit $\eta$ defined as $\eta = R_{\text{UMR}}/(R_0 j B)$ in Ref.[45], where in our case, due to the different symmetries of Te, we use $|A^{\text{diff}}|/2$ as $R_{\text{UMR}}$ and $R(B = 0\text{ T})$ as $R_0$, while $j$ is the current density ($j \sim 10^7$ A/m$^2$, see Supplementary Section 1) and $B$ the magnetic field applied. As a result, we obtained an $\eta \sim 5 \times 10^{-6}$ cm$^2$/(AT), which is one order of magnitude larger than the highest value reported for Ge(111) at 15 K ($\eta \sim 4.2 \times 10^{-7}$ cm$^2$/(AT)) in ref.[45] and at least three orders of magnitude larger than observed in SrTiO$_3$ (ref.[50]), Bi$_2$Se$_3$ (ref.[44]) and α-GeTe (ref.[11]). The large UMR magnitude can be explained by the strong Edelstein effect, as detailed below.

Finally, we focus on the control of the UMR by a gate voltage ($V_G$). We will show how the shifts in energy of the Fermi level ($\varepsilon_F$) of the Te NWs leads to the tuning the Edelstein effect. In Fig. 4a, we show the magnetoresponse evolution under changes of carrier density, driven by $V_G$ modulation. To account for the variation of resistance with the gate (see Supplementary Section 7), we normalize $R^{\text{diff}}(\alpha)$ measured at different $V_G$ with respect to the value $R^{\text{avg}}(90°)$. For negative $V_G$, i.e., lowering the Fermi level and increasing the hole concentration, the magnetoresponse trend is relatively flat except for the two prominent features at $\alpha = 0°$ and $\alpha = 180°$, indicating that the electrical current induces a significant spin polarization only in the z-direction. Increasing $V_G$, i.e., decreasing the hole concentration and bringing the Fermi level closer to the band gap, we observe an increase in the amplitude of the UMR (up to a factor of 6, see Fig. 4b) and the appearance of shoulders close to the peaks at $\alpha = 0°$ and $\alpha = 180°$. Similar shoulders were observed for $V_G = 0$ V in some NWs, including those shown in Fig. 3f-i. Their relative intensity was found to change from sample to sample (see Supplementary Section 8), probably due to the slightly different doping in different NWs.

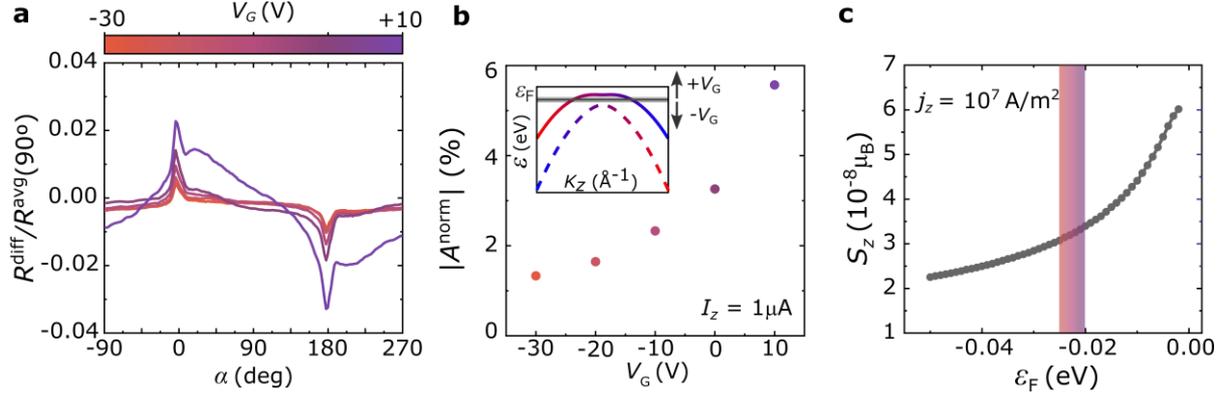

**Fig. 4| Gate modulation of the unidirectional magnetoresistance and comparison with theory. a**, Normalized $R^{\text{diff}}$ angular dependences measured at $B = 9$ T, $T = 10$ K, and $\pm I_z = \pm 1$ µA, for a left-handed Te NW applying different gate voltages $V_G$. **b**, Gate voltage dependence of the normalized signals' amplitude $|A^{\text{norm}}| = [R^{\text{diff}}(0°) – R^{\text{diff}}(180°)]/R^{\text{avg}}(90°)$, in inset is shown the shift of $\varepsilon_F$ induced by the gate with respect to the Te band structure (the colours display the spins projection to the z-axis). **c**, Spin per unit cell ($S_z$) induced by a current density $j_z = 10^7$ A/m$^2$ through the Edelstein effect for different Fermi energy positions. The zero energy corresponds approximately to the valence band edge, and the highlighted zone represents the energy range corresponding to the applied gate voltages.

The increase of the magnetoresponse at positive $V_G$ can be correlated with the Te band structure. Looking at the band structure at different energies, a change of the Fermi contour in k-space is expected to cause a variation of the Edelstein effect efficiency[51] and, consequently, a variation of the magnetoresponse. In Fig. 4b,c, we compare the gate dependence of $|A^{\text{norm}}| = [R^{\text{diff}}(0°) – R^{\text{diff}}(180°)]/R^{\text{avg}}(90°)$ with the calculated spins per unit cell ($S_z$), generated by the Edelstein effect at different $\varepsilon_F$ and a constant charge current density of $10^7$ A/m$^2$ (see Methods).

In order to compare experiments and calculations, we measured the carrier density of a Te NW through ordinary Hall measurements to estimate the position of $\varepsilon_F$ in the band structure shown in Fig. 1e. The extracted hole density was $p = 7.4 \times 10^{17}$ cm$^{-3}$ at $V_G = 0$ V and it could be varied in the range 7-8$\times 10^{17}$ cm$^{-3}$ using the back-gate voltage ranging from -30 V to 10 V (see Supplementary Section 9). Interestingly, the rather small modulation in $p$ induces significant changes on the recorded signals, highlighting that it is possible to finely tune the Edelstein effect in Te by acting on its band structure through standard electrostatic gating.

Our calculations on the charge carrier density reveal that the extracted $p$ corresponds to an $\varepsilon_F$ approximately 20 meV below the band edge, which can be tuned in a 5 meV energy window by gating. Figure 4c shows that $S_z$, which is associated to the efficiency of the Edelstein effect creating the spin polarisation, increases when $\varepsilon_F$ is moved towards the valence band edge, in good agreement with the increase of $|A^{norm}|$ with $V_G$ displayed in Fig. 4b. Moreover, we highlight that the calculated $\varepsilon_F$ position implies a single band occupancy, since the band which lies closest in energy is populated at energies 130 meV below the valence band edge (see Supplementary Section 9). This situation is ideal to maximize the Edelstein effect and thus the UMR, since the occupation of the lower band, with opposite radial spin texture, would partially compensate the induced spin density (see Supplementary Section 10 and 11 for more details on the calculations and for the full $S_z$ vs $\varepsilon_F$).

Conclusion

In conclusion, we demonstrated all-electrical generation, manipulation and detection of chirality dependent spin polarization in single crystalline Te NWs. The spin polarization from chiral origin gives rise to a UMR that is one-to-several orders of magnitude larger than the reported in other non-chiral systems. The UMR is also tuneable with electrical gating, providing us with an extra knob for controlling the spin polarization. These effects are induced by an Edelstein effect emerging from the Te radial spin texture. Unlike conventional Rashba systems and similarly to the CISS observed in organic molecules, the induced spin polarization is oriented along the current direction. We think that our description this phenomenon on the basis of the Edelstein effect might be extended to other chiral systems characterized by translational symmetry. Our results put on a firm ground the fundamental interplay between structural chirality and electron spin in inorganic chiral Te NWs, enabling the design of unconventional all-electrical devices in which the spin polarization is not determined by the direction of a magnetization, but by the handedness of the crystal.

Methods

**Chemical synthesis of Te NWs**

The synthesis consists of a high-temperature reduction of a tellurium oxide in presence of hydrazine ($N_2H_4$) in a basic aqueous medium[35–37,52]. We followed the hydrothermal growth recipes proposed in literature[36,37]. Briefly, we mixed $Na_2TeO_3$ (104 mg) and polyvinylpyrrolidone (average $M_w$ 360,000 - PVP360, 547.9 mg) in 33 mL of MilliQ® water by magnetic stirring up to achieve a clear solution at room temperature. Then, $NH_4OH$ solution (3.65 mL, 25%w in water) and hydrazine hydrate (1.94 mL, 80%, w/w%) were added while stirring. The mixture is transferred to an autoclave that was sealed and heated at 180°C for 23 h. We washed the resulting material by successive centrifuge-assisted precipitation (5000 rpm - 5 min) and redispersion with MilliQ® water (10 steps of 4 mL water each). At this stage, we are still not able to produce selectively one or the other enantiomer and the distribution of left- and right-handed NWs is random.

**Sample preparation**

Te NWs were redispersed in a dimethylformamide: $CHCl_3$ mixture (1.3:1 vol.) to be used for the drop casting of solution droplets at the deionized water-air interface in homemade Langmuir trough. After the evaporation of the solvent, Te NWs floating on the water surface were picked-up (Langmuir-Schaefer technique) with Si/$SiO_2$ substrates (Si doped n+, 5×5 mm, 300 nm thermal oxide). Isolated NWs with opportune dimensions were selected with an optical microscope, without knowing a priori the handedness of the NWs. The contacts were defined through e-beam lithography performed on poly(methylmethacrylate) (PMMA) A4/PMMA A2 double layer and then Pt was deposited by sputtering. Seven NWs based devices (D1-D7) were fabricated and used for magnetotransport measurements at low temperature. Results obtained from D1 are shown in Fig. 1d (left-handed), Fig. 2 b,c,d and Fig. 3 d,f,h. Results obtained from D2 are shown in Fig. 4 a,b. Results obtained from D3 are shown in Fig. 1b,d (right-handed), Fig. 2a and Fig. 3 e,g,i. Devices D4-D7 are included in the supporting information.

**TEM imaging**

Two devices showing opposite magneto responses were sectioned by FIB for STEM chirality characterisation. Two lamellae were prepared from every device: one perpendicular to the wire in the contact are (to validate the quality of electrical contact) and one along the wire for measuring the chirality (see Supplementary Section 1). Helios 600 DualBeam™ (ThermoFisher, USA) was used for lamellae preparation and SEM imaging, TitanG2 60-300 operated at 300kV without STEM corrector was used for STEM imaging. For chirality determination samples were imaged in either of <010> zones and then tilted ±30º resulting in mirrored images. The handedness was determined by comparison to atomic models of the known chirality. The absence of the mirror plane between the sample and digital image has been ruled out by imaging alphanumeric grid. The STEM analysis was performed after the samples were already measured electrically.

**Magneto transport measurements**

The devices were wire-bonded to a sample holder and installed in a physical property measurement system (PPMS, Quantum Design) for transport measurements with a temperature range of 2–400 K and maximum magnetic field of 9 T. We performed measurements of d.c. longitudinal and transversal resistance using a Keithley 6221 current source and a Keihtley 2128 nanovoltmeter. Asymeric ($I+/0$ and $I-/0$) and symmetric ($I+/I-$) delta mode (16 to 32 counts) were employed to improve the signal to noise ratio (with current ranging from 50 nA to 10 μA). The average of the signals obtained from the asymmetric delta mode measurements was equivalent to the signals obtained from the symmetric delta mode and corresponds to the current-independent resistance ($R^{avg}$). The half difference of the signals obtained from the asymmetric delta mode was taken to analyse the current-dependent resistance ($R^{diff}$). For the angle dependent measurements, the magnetic field was rotated in the plane of the sample. In the current dependent angular dependences, we excluded the current that were high enough to raise the local temperature of the NWs. For the gate dependence measurements, a Keithley 2636 source meter was used to apply a constant voltage to Si doped substrate, while monitoring the leakage current through the $SiO_2$ dielectric to be smaller than 10 nA.

**Calculations on band structure and Edelstein effect**

The states relevant for electric transport are located around the H point. Therefore, we use an effective model Hamiltonian which reproduces the band effective band structure introduced in Ref.[39] as well as the radial spin texture[46,47] to model the valence bands around the H point,

$$H(\mathbf{k}) = -\Delta - Ak_z^2 - B\sqrt{k_x^2 + k_y^2} + \chi\left[\frac{\Delta}{\sqrt{k_x^2 + k_y^2}}(k_x\hat{\sigma}_x + k_y\hat{\sigma}_y) + \beta k_z\hat{\sigma}_z\right] \quad (1)$$

Here, $k_z$ is along the H-K direction and the Pauli spin vector $\hat{\boldsymbol{\sigma}}$ represents the spin degree of freedom. We use the parameters proposed in Ref. 23, $\Delta$=63 meV, $A$=3.64 × 10$^{-19}$ eV m$^2$, $B$ =3.26 × 10$^{-19}$ eV m$^2$, $\beta$ = 2.4 × 10$^{-10}$ eV m. $\chi$ corresponds to the chirality with $\chi = 1$ (left-handed) and $\chi = -1$ (right-handed), respectively. The corresponding effective band structure is shown in Fig. 1e, the spin texture in Fig. 1f.

We calculate the charge current density $\mathbf{j}$ as well as the spin density $\mathbf{S}$ response to an external electric field $\mathbf{E}$,

$$\mathbf{j} = -\frac{e}{V}\sum_{\mathbf{k}} \mathbf{v_k} f_\mathbf{k} \quad (2)$$

$$\mathbf{S} = \sum_{\mathbf{k}} \langle \mathbf{s} \rangle_\mathbf{k} f_\mathbf{k} \quad (3)$$

Here, e is the absolute value of the elementary charge, $V$ is the sample volume, $\mathbf{v_k} = \partial\varepsilon/\hbar\partial\mathbf{k}$ is the group velocity, $\langle\mathbf{s}\rangle_\mathbf{k}$ is the spin expectation value, $\varepsilon$ the energy dispersion, and $f_\mathbf{k}$ is the distribution function. Its evolution under the influence of an external electric field $\mathbf{E}$ as well as scattering is described by the Boltzmann equation,

$$-\frac{e}{\hbar}\mathbf{E}\frac{\partial f_\mathbf{k}}{\partial \mathbf{k}} = \left.\frac{\partial f_\mathbf{k}}{\partial t}\right|_{\text{scatt}} \quad (4)$$

where we have assumed the system to be stationary and spatially homogeneous. $\partial f_\mathbf{k}/\partial t|_{\text{scatt}}$ is the scattering term which we approximate by the relaxation time approximation

$$\left.\frac{\partial f_\mathbf{k}}{\partial t}\right|_{\text{scatt}} = -\frac{1}{\tau_\mathbf{k}} g_\mathbf{k} \quad (5)$$

Here $g_\mathbf{k} = f_\mathbf{k} - f_\mathbf{k}^0$ is the nonequilibrium part of the distribution function, $f_\mathbf{k}^0$ is the Fermi Dirac distribution function and $\tau_\mathbf{k}$ is the momentum relaxation time. The Boltzmann equation (4) is solved by

$$f_\mathbf{k} = f_\mathbf{k}^0 + \frac{\partial f_\mathbf{k}^0}{\partial \varepsilon} e \tau_\mathbf{k} \mathbf{v}_\mathbf{k} \cdot \mathbf{E} \qquad (6)$$

Thus, the electric field **E** leads to a reoccupation of states, depending on their group velocity and momentum relaxation time. In a simplified picture, this can be understood as a "shift" of the Fermi contour in **k** space. Inserting Eq.(6) into Eqs. (2) and (3) yields

$$\mathbf{j}_c = -\frac{e^2}{V} \sum_\mathbf{k} \frac{\partial f_\mathbf{k}}{\partial \varepsilon} \tau_k \mathbf{v}_\mathbf{k} (\mathbf{v}_\mathbf{k} \cdot \mathbf{E}) =: \sigma \mathbf{E} \qquad (7)$$

$$\mathbf{S} = e \sum_\mathbf{k} \frac{\partial f_\mathbf{k}}{\partial \varepsilon} \tau_k \langle \mathbf{s} \rangle_\mathbf{k} (\mathbf{v}_\mathbf{k} \cdot \mathbf{E}) =: \kappa \mathbf{E} \qquad (8)$$

In our calculations, we assume $T = 0$ and a constant relaxation time $\tau_0$. Hence, only states at the Fermi level contribute to the transport. Using Eqs. (7) and (8) we calculate the charge current as well as the current-induced spin density in bulk Te, which is shown as a function of the Fermi level in Fig. 4c. To calculate the induced spin density, we use $S_z = \kappa_{zz} j_z / \sigma_{zz}$, where **j** and σ are the current density and the charge conductivity respectively, and the Edelstein susceptibility κ is the tensor which connects the induced spin density **S** to the applied electric field, $\mathbf{S} = \kappa \mathbf{E}$ (see supplementary section 10 for more details).

Acknowledgements

This work is supported by the Spanish MICINN under Projects RTI2018-094861-B-I00 and PID2019-108153GA-I00 and under the Maria de Maeztu Units of Excellence Programme (MDM-2016-0618), by the European Union H2020 under the Marie Slodowska Curie Actions (0766025-QuESTech and 892983-SPECTER), and by Intel Corporation under 'FEINMAN' and 'VALLEYTRONICS' Intel Science Technology Centers. B.M.-G. acknowledges support from Gipuzkoa Council (Spain) in the frame of Gipuzkoa Fellows Program. M.S.R acknowledges support from "la Caixa" Foundation (ID 100010434) with code LCF/BQ/DR21/11880030. M.G. acknowledges support from la Caixa Foundation (ID 100010434), for a Junior Leader fellowship (Grant No. LCF/BQ/PI19/11690017). A.J. acknowledges support from CRC/TRR 227 of Deutsche Forschungsgemeinschaft (DFG).

Author contribution

F.Calavalle, M.G., F.Casanova and L.E.H. conceived the study. F. Calavalle and M.S.-R. fabricated the samples and performed the magnetotransport measurements with the help of D.C.V. and H.Y. B.M.-G. synthetized the Te NWs with the support of A.M-A. and A.C. performed the STEM analysis. A.J. conducted the theoretical calculations with the support of I.M. I.V.M. and S.O. performed the ab-initio calculations. F.Calavalle and M.G. wrote the manuscript with input from all authors. All authors contributed to the discussion of the results and their interpretation.

Competing interests

The authors declare no competing financial interest.

pyrrolidone)-assisted hydrothermal process. *Langmuir* **23**, 10873 (2007).